\documentclass[11pt]{article}
\usepackage{geometry}                

\usepackage{slashed}

\usepackage{jheppub}
\usepackage{bbm}
\usepackage{bbold}

\usepackage{amsmath,amssymb,graphicx} 

\usepackage{psfrag}
\usepackage{subfigure}
\usepackage{tabularx}
\usepackage{footmisc}

\newcommand\cc[1]{#1^{^{\kern-6pt \circ}}\kern2pt}

\newcommand{\F}{{\cal F}}
\newcommand{\dd}{{\rm d}}

\def\pa{\partial}
\renewcommand{\a}{\alpha}

\renewcommand{\d}{\delta}

\newcommand{\m}{\mu}
\newcommand{\n}{\nu}

\def\be{\begin{equation}}
\def\ee{\end{equation}}
\def\bea{\begin{eqnarray}}
\def\eea{\end{eqnarray}}
\def\ba{\begin{array}}
\def\ea{\end{array}}
\def\bi{\begin{itemize}}
\def\ei{\end{itemize}}
\def\half{{\textstyle{1\over2}}}

\newcommand{\beq}{\begin{equation}}
\newcommand{\eeq}{\end{equation}}
\newcommand{\beqn}{\begin{eqnarray}}
\newcommand{\eeqn}{\end{eqnarray}}
\newcommand{\bga}{\begin{align}}

\def\dalemb#1#2{{\vbox{\hrule height .#2pt
\hbox{\vrule width.#2pt height#1pt \kern#1pt
\vrule width.#2pt}
\hrule height.#2pt}}}

\def\a{\alpha}

\def\g{\gamma}
\def\d{\delta}

\def\t{\tau}

\def\dd{\mbox{d}}

\def\O{\Omega}

\def\a{\alpha}

\def\d{\delta}

\def\g{\gamma}

\def\e{\epsilon}
\def\ve{\varepsilon}

\def\f{\phi}
\def\F{\Phi}

\def\k{\kappa}
\def\l{\lambda}
\def\L{\Lambda}
\def\m{\mu}
\def\n{\nu}

\def\t{\tau}

\def\pa{\partial}

\newcommand{\tn}[1]{\mbox{\tiny #1}}
\renewcommand{\@}[1]{\sqrt{#1}}
\renewcommand{\le}[1]{\label{#1}\end{eqnarray}}

\newcommand{\eq}[1]{(\ref{#1})}
\def\nn{\nonumber\\}

\def\half{{1\over2}\,}

\title{\LARGE Instantons and the Hartle-Hawking-Maldacena Proposal for dS/CFT}
\author[a]{\bf Sebastian de Haro} 
\author[b]{and \bf Anastasios C. Petkou}
\affiliation[a]{Institute for Theoretical Physics and
Amsterdam University College \\
 University of Amsterdam
\\Science Park 113, 1098 XG Amsterdam, The Netherlands}
\affiliation[b]{Institute of Theoretical Physics\\ Aristotle University of Thessaloniki \\54124 71003 Thessaloniki, Greece}

\emailAdd{s.deharo@uva.nl}
\emailAdd{petkou@physics.auth.gr}

\date{\today}                                          

\begin{document}

\abstract{We test the Maldacena proposal for the Hartle-Hawking late time quantum state in an asymptotically de Sitter universe. In particular, we calculate the on-shell action for scalar instantons on the southern hemisphere of the four-sphere and compare the result with the renormalized on-shell action for scalar instantons in EAdS$_4$. The two results agree provided the corresponding instanton moduli as well as the curvature radii are analytically continued.
The instanton solutions in de Sitter are novel and satisfy mixed boundary conditions. We also point out that instantons on $S^4$ calculate the regularized volume of EAdS$_4$, while instantons on EAdS$_4$ calculate the  volume of $S^4$, where the boundary condition of the instanton in one space is identified with the radius of curvature of the other. We briefly discuss the implications of the above geometric property of instantons for higher-spin holography.}
\maketitle

\section{Introduction}

The task of extending the holographic principle to an explicitly time-dependent, cosmological, setting proves to be as non-trivial as it is interesting. A particularly simple approach is Maldacena's proposal \cite{Maldacena1} (see also \cite{Harlow}) for the evaluation of the Harte-Hawking (HH) late time quantum state in an asymptotically de Sitter universe\footnote{Important earlier attempts for dS/CFT include \cite{Strominger,Verlinde,Balasubramanian1,Balasubramanian2}.}. The HH state is specified \cite{HH} by a Euclidean wave function of the schematic form:
\bea\label{HHwf}
\Psi_{\tn{HH}}[\chi]=\int_{\phi|_{\partial M}\equiv \chi} {\cal D}\phi~e^{-S[\phi]}~,
\eea
where ${ \partial M}$ is a 3-dimensional spacelike hypersurface near the future infinity of an asymptotically de Sitter space-time with radius $\ell_{\tn{dS}}$. We have collectivelly denoted the bulk fields as $\f$. Then, Maldacena's proposal entails that the HH state can be obtained if one calculates the corresponding renormalized on-shell action on Euclidean AdS$_4$ (EAdS$_4$) with radius $\ell_{\tn{AdS}}$,  and performs the analytic continuation $\ell_{\tn{AdS}}\rightarrow i\ell_{\tn{dS}}$. Since the EAdS$_4$ on-shell action is reasonably well defined---it gives the partition function of a Euclidean 3-dimensional CFT---Maldacena's proposal gives a way to make sense of and calculate the HH wave function from AdS/CFT. When analytically continued to Lorentzian signature, \eq{HHwf} gives the Bunch-Davies wave function in de Sitter space.

AdS/CFT works best when there is an explicit string theory realisation of the bulk physics. At present there is no satisfactory string theory description for gravity with a positive cosmological constant, but an alternative and more adventurous set up for a concrete realisation of dS/CFT was proposed in \cite{AHS}. This is based on Vassiliev's higher-spin (HS) theory\footnote{For a recent review of HS theories see \cite{Misha}.} which provides the only known consistent classical description of interacting higher-spin gauge fields in a de Sitter background. Its suggested holographic dual is the Euclidean $\mbox{Sp}(N)$ vector model with anti-commuting scalars and it is a free CFT$_3$ when the higher-spin symmetry is unbroken. Since the quantum properties of free theories are known, the HS version of dS/CFT offers some hope of understanding quantum gravity (plus HSs) on dS$_4$. Moreover, an analytic continuation is also at work here, as  $N\mapsto-N$, where $N\sim{\ell^2_{\tn{dS}}/ G_{\tn{N}}}$, maps the $\mbox{Sp}(N)$ anti-commuting vector model to usual (commuting) $\mbox{O}(N)$ vector model which is believed to be the holographic dual of HS theory on AdS$_4$. Notice that Newton's constant $G_{\tn{N}}$ is held fixed.

The Hartle-Hawking-Maldacena (`HHM') proposal has been tested  in a number of cases. The most explicit check in \cite{Maldacena2} involves pure gravity only: the on-shell action on Euclidean de Sitter (the four-sphere) is shown to analytically continue to the finite part of the volume of EAdS$_4$, after dropping counterterms which are argued to give only imaginary phases which do not contribute to the measure $|\Psi_{\tn{HH}}|^2$ \cite{Maldacena1,Maldacena2}. It has also been tested in the presence of scalars and gauge fields by showing that their generating functionals for two-point functions in EAdS$_4$  and dS$_4$ are related by suitable analytic continuations \cite{Harlow,Castro,Anninos:2014hia}. A discussion of the HHM proposal and the relevant analytic continuation for cosmological theories with matter fields and generic potentials would entail a discussion of exact non-trivial solutions which come with a moduli space, and then one wonders how the latter transforms under the analytic continuation in the HHM proposal. Also, it is important to know how boundary deformations (such as multiple trace deformations) of the CFT dual to EAdS$_4$ carry over to de Sitter space, and whether some complexification of the deformation parameter is involved. Finally, one may ask whether there exists an analytic continuation at the level of fields rather than particular solutions. For recent important progress, see \cite{kostasmcfadden}, where it was shown that holography correctly reproduces both the spectra and the non-gaussianities for general inflationary space-times, i.e.\ for any potential that supports inflationary FRW solutions.

In this short note we study exact solutions in a scalar theory conformally coupled to gravity and check that the HHM proposal works, namely the HH state is given by an analytic continuation of the holographic EAdS$_4$ partition function. These solutions are instantons: they are zero energy, exact solutions of the Euclidean equations of motion with finite action. Such solutions for scalar fields exist in EAdS$_4$ \cite{SdHTP,dHPP} and, as we show here, also in 4-dimensional de Sitter space. Further investigation of these solutions is relegated to a companion paper \cite{next}. We test the HHM proposal for these solutions and  find exact agreement, provided the EAdS$_4$ radius as well as the moduli of the solution and their boundary conditions are analytically continued. 
We also note there exists a simple geometric description of our results. In particular, the renormalized on-shell action of scalar instantons on EAdS$_4$ evaluates the volume of the four-sphere, while the corresponding on-shell action of instantons on $S^4$ evaluates the regularized volume of EAdS$_4$. In both cases, the instanton moduli serve as regulators of the corresponding volume forms. The above geometric description allows us to interpret the on-shell action of scalar instantons as the free-energy of a theory on $S^3$ and this in turn may have  implications for the holography of HS theories. 

\section{Instanton Solutions in Euclidean AdS$_4$ and dS$_4$}

\subsection{The HH state and holographic partition functions} 

A central object for the holographic principle is the properly renormalized bulk on-shell action. This can be evaluated as a functional of the boundary conditions for a suitably regular solution of the bulk equations of motion. In this case, depending on the choice of boundary conditions it yields either a generating functional for quantum correlation functions or an effective action for a putative CFT living on the boundary.   

However, given a {\it particular} solution of the bulk equations of motion, namely one where the bulk fields assume fixed boundary values, the renormalized bulk on-shell action evaluates the {\it free energy} of the boundary theory and therefore the partition function as:
\be
Z=e^{-F}\equiv e^{-S_{\tn{on-shell}}}~.
\ee
For example, the renormalized Einstein-Hilbert action using the Poincar\'e patch of EAdS$_4$ gives zero, while the corresponding quantity evaluated using a global parametrization of the bulk metric with conformal boundary metric $S^3$  yields the non-zero free energy of a CFT on $S^3$. 

In the presence of bulk matter fields, things are less clear as exact solutions of the corresponding nonlinear equations of motion are uncommon. Notable exceptions are instantons in EAdS$_4$ \cite{SdHTP, dHPP}. In particular, as we we review in Appendix A, for conformally coupled scalars in EAdS$_4$ with boundary behaviour $\phi(z,\vec{x})\rightarrow z\,\phi_{(0)}(\vec{x})+z^2\,\phi_{(1)}(\vec{x})$, where $\phi_{(0)}(\vec{x})$ takes a {\it fixed} form, the bulk on-shell action $S_{\tn{on-shell}}[\phi_{(0)}]=-\ln \tilde{Z}_0$ gives the logarithm of the partition function $\tilde{Z}_0$ of the {\it dual} boundary CFT, namely the CFT having in its spectrum the operator ${\cal O}_1$ of dimension $\Delta=1$ and $\langle{\cal O}_1\rangle\sim \phi_{(0)}$. Moreover, having this result one can calculate by a Legendre transform the partition function of the usual boundary CFT, namely the one having an operator ${\cal O}_2$ of dimension $\Delta=2$ and $\langle{\cal O}_2\rangle\sim \phi_{(1)}$. In \cite{dHPP} the partition function $\tilde{Z}_0$ was interpreted as giving the probability for the nucleation of the instanton vacuum in the boundary theory. 

Regarding now the HH state for an asymptotically dS$_4$ universe, a simple example arises in the absence of matter fields when it is given by the on-shell value of the EH action with a positive cosmological constant. After the analytic continuation of dS$_4$ to the 4-sphere, the result is finite and proportional to the volume of $S^4$. It is then easily seen \cite{Maldacena2} that this is analytically continued to the finite part of the holographic partition function on $S^3$.  As in the EAdS$_4$ case, non-trivial results for the HH state in the presence of bulk matter are uncommon. In this note we will improve on this situation by providing results for dS$_4$ instantons which, as we will see, are intimately related to the usual EAdS$_4$ ones. Our main aim is therefore to evaluate both the HH state and the holographic partition functions of systems involving gravity and matter fields and to test whether they are still related by analytic continuation. In doing so, we will obtain some interesting new results. 

\subsection{Action and solutions}

In the context of AdS/CFT there are few examples where particular exact solutions of the bulk equations of motion of gravity coupled to scalars are known. Exact solutions are needed in order to calculate the exact partition function of the CFT. A non-trivial example was studied in \cite{SdHTP,dHPP}. In this paper we write down similar solutions for de Sitter space and use them to test the HHM proposal.

The Euclidean bulk action we will consider contains the Einstein-Hilbert term, a conformally coupled scalar with a $\phi^4$ potential, and the Gibbons-Hawking term modified by a coupling to the conformal scalar. It should be noted from the outset that for $\Lambda<0$ and the special value  of the quadratic coupling $\lambda=\lambda_{\tn{cr}}=2\kappa^2/\ell_{\tn{AdS}}^2$, this action is a consistent truncation of ${\cal N}=8$ sugra to the diagonal of the Cartan subgroup $\mbox{U}(1)^4$ of the $\mbox{SO}(8)$ gauge group \cite{MTZ,dHPP}. We will encounter this special value of $\lambda$ later on. 

As usual, we regulate the theory introducing a large distance cutoff $\e$ which we send to zero at the end of the calculation \cite{SdHKSSS}\footnote{See also \cite{kostas}.}:
\bea\label{action}
S&=&{1\over2\k^2}\int_{M_\e}\dd^4x\,\sqrt{g}\left(-R+2\L\right)+\int_{M_\e}\dd^4x\,\sqrt{g}\left(\half(\pa_\m\f)^2+{R\over12}\,\f^2+{\l\over4!}\,\f^4\right)\nn
&-&{1\over2\k^2}\int_{\pa M_\e}\dd^3x\,\sqrt{\g}~2K\left(1-{\k^2\over6}\,\f^2\right),
\eea
where $\k^2=8\pi G_{\tn{N}}$ and the cosmological constant can be either positive or negative.
The difference between a positive and a negative cosmological constant $\L=\mp{3\over\ell^2}$ is in the location and orientation of the regulated conformal boundary $\pa M_\e$ and in the counterterms:
\bea
S_{\tn{ct}}^{\tn{EAdS}}&=&{1\over\k^2}\int_{\pa M_\e}\dd^3x\,\sqrt{\g}\left({4\over\ell}+\ell\,R[\g]\right)~.
\eea
For the Euclidean dS case, no counterterms are needed because the wave function \eq{HHwf} only includes contributions from configurations that are asymptotically regular.

The instanton solutions are constructed using the Weyl invariance of the matter part of the action, hence it is useful to use global coordinates that are conformal to $I\times\pa M$, where $I$ is a (finite or infinite) interval. In the cases at hand $\pa M=S^3$. In EAdS$_4$ we will use conformal cylinder coordinates:
\bea
\dd s^2_{\tn{EAdS}}={1\over\sinh^2{\t\over\ell}}\left(\dd\t^2+\ell_{\tn{AdS}}^2\,\dd\O_3^2\right),~~\t\in(0,\infty)~.
\eea
For the purposes of the HHM proposal we will work with half\footnote{Since our scalar instantons do not back-react, the calculation of the HH wave functional involves gluing the Euclidean half $S^4$ in the past to Lorentzian dS$_4$  in the future. Since we are interested in the norm of the HH wave functional,  considering half $S^4$ suffices as the configuration along the imaginary (i.e. Lorentzian) path just gives a purely imaginary  contribution to the on-shell action.} Euclidean de Sitter, i.e.~the southern hemisphere of $S^4$:
\bea
\dd s^2_{S_-^4}={1\over\cosh^2{r\over\ell_{\tn{dS}}}}\left(\dd r^2+\ell_{\tn{dS}}^2\,\dd\O_3^2\right),~~r\in(-\infty,0]~.
\eea
Lorentzian de Sitter is obtained by Wick rotating back $r\rightarrow-i\ell_{\tn{dS}}\,\t$. 

The relevant solutions of the scalar sector \cite{next} are obtained by solving the Klein-Gordon equation:
\bea\label{KG}
\Box\f-{2\L\over3}\,\f-{\l\over6}\,\f^3=0
\eea
together with the requirement that the stress-energy tensor vanishes. The latter requirement turns out to give \cite{Iannis}:
\bea
\left(\nabla_\m\nabla_\n-{1\over4}\,g_{\m\n}\Box\right)\f^{-1}=0~.
\eea
The solutions on EAdS$_4$ and $S^4$ are then given by:
\bea\label{solutions}
\f_{\tn{EAdS}_4}^{\ve,b_I}(\t,\O_3)&=&{\ve\,\sinh{\t\over\ell}\over b_0\cosh{\t\over\ell} +b_5\sinh{\t\over\ell}+b_i\O_i} \label{CCCsolution}\nn
\f^{\ve,a_I}_{S^4}(r,\Omega_3)&=&{\ve\,\cosh{r\over\ell}\over a_0\sinh{r\over\ell} +a_5\cosh{r\over\ell}+a_i\O_i} \label{S4solution}~,
\eea
where $i=1,\ldots,4$ and $\O_i$ is a unit vector normal to the three sphere. $\ve$ can take the values $\ve=\pm1$. The requirement that these are solutions to the equation of motion \eq{KG} renders the moduli space  non-trivial,
\bea\label{modulispacesAdS}
\mbox{on EAdS}_4:~~~-b_0^2+b_5^2+b_i^2&=&{\l_{\tn{AdS}}\over12}\,\ell_{\tn{AdS}}^2~,~~i=(1,\ldots,4)\\
\label{modulispacesdS}
\mbox{on}~~S^4:~~~~~~a_0^2-a_5^2+a_i^2&=&{\l_{S^4}\over12}\,\ell_{\tn{dS}}^2~.
\eea
From (\ref{modulispacesAdS}) and (\ref{modulispacesdS})  we see that the moduli spaces of instantons on EAdS$_4$ and dS$_4$ are themselves EAdS$_4$ or dS$_4$ depending on the sign and values of the quartic couplings $\lambda_{\tn{AdS}}$ and $\lambda_{\tn{dS}}$; more specifically, the curvatures depend on the particular combinations ${\l_{\tn{AdS}}\over12}\,\ell^2_{\tn{AdS}}-b_5^2$ and ${\l_{S^4}\over12}\,\ell_{\tn{dS}}^2-a_0^2$. In the next subsection we will see the reason for this: $b_5$ and $a_0$ are not moduli of the solution; they parametrize boundary conditions instead. From the boundary point of view, $b_5$ is the marginal coupling of a triple-trace deformation of the CFT. Therefore the moduli space of the solutions is EAdS$_4$ if ${\lambda_{\tn{AdS}}\over12}\,\ell^2_{\tn{AdS}}<b_5^2$, 
which is the condition required for regularity of the bulk solution. At the critical value ${\lambda_{\tn{AdS}}\over12}=b_5^2$ the effective potential of the dual field theory becomes unbounded from below, which was interpreted as an instability of the dual theory against marginal deformations, decaying via quantum tunneling \cite{dHPP}.

\subsection{Boundary conditions}

Not all of the parameters $a_I,b_I$ ($I=0,\ldots,5$) are moduli as we now show. Two of them are boundary conditions relating the leading and subleading modes of the scalar: 
\bea\label{bcs}
\f_{S^4_-}(-\e\,\ell,\Omega)&=&{\ve\over a_5+a_i\O_i}+{\e\,\ve\, a_0\over(a_5+a_i\O_i)^2}+{\cal O}(\e^2) =\F_{(0)}^{a_5,a_i} +\e\,\F_{(1)}^{a_5,a_i}+{\cal O}(\e^2)\\
\f_{\tn{EAdS}}(\e\,\ell,\O)&=&{\e\,\ve\over b_0+b_i\O_i}-{\e^2\,\ve\, b_5\over(b_0+b_i\O_i)^2}+{\cal O}(\e^3)=\e\,\F_{(0)}^{b_0,b_i}+\e^2\,\F_{(1)}^{b_0,b_i}+{\cal O}(\e^3)~.\nonumber
\eea
The leading and subleading terms in the expansion of the field are related by:
\bea\label{bdydef}
\F_{(0)}^{a,a_i}(\O)&\equiv&{\ve\over a+a_i\O_i}\nn
\F_{(1)}^{a,a_i}(\O)&=&\pm\,\ve\,\a\left(\F^{a,a_i}_{(0)}(\O)\right)^2~,~~\mbox{EAdS}/S^4_-~,
\eea
where $\a=b_5$ for EAdS$_4$ and $\a=a_0$ for $S_-^4$. Thus, $b_5,a_0$ parametrize  marginal triple trace deformations that change the boundary conditions from Dirichlet to mixed \cite{dHPP} (see also \cite{Iannis}).

\subsection{On-shell action and the HH wave function}

We stress that, given the action \eq{action}, the solutions \eq{solutions} are exact: they have zero stress-energy tensor hence the EAdS/dS background stays unmodified. Thus we can compute the on-shell effective action including its finite part\footnote{As mentioned before,  the EAdS$_4$ solutions can be embedded in M-theory. We are not considering $1/N$ corrections here, but to the given order in the bulk coupling we can trust the result for the finite part of the on-shell action including its $\l$-dependence, and this is only possible because the solution is exact.}. For simplicity we set the spherical modes $a_i=b_i=0$ and get:
\bea\label{finalEAdS}
S^{\tn{on-shell}}_{\tn{EAdS}}&=&{4\pi^2\ell^2_{\tn{AdS}}\over\k^2}-{\l_{\tn{AdS}}\,\pi^2\ell^4_{\tn{AdS}}\over12}{2b_0+b_5\over3b_0^3(b_0+b_5)^2}+\pi^2\ell^2_{\tn{AdS}}~{b_5\over b_0^3}+{\cal O}(\e)\nn
&=&{4\pi^2\ell^2_{\tn{AdS}}\over\k^2}+{2\pi^2\ell^2_{\tn{AdS}}\over3b_0^3}{b_0^2+b_0b_5+b_5^2\over b_0+b_5}+{\cal O}(\e),
\eea
where in the last line we used \eq{modulispacesAdS}. In the special case of a Neumann boundary condition $b_5=0$, which requires negative coupling $\l_{\tn{AdS}}<0$, we reproduce the result in \cite{SdHTP},
\bea\label{zerocase}
S_{\tn{on-shell, EAdS}}^{b_5=b_i=0}={4\pi^2\ell^2_{\tn{AdS}}\over\k^2}-{8\pi^2\over\l_{\tn{AdS}}}~.
\eea
Notice that this is a positive quantity. The case $\l_{\tn{AdS}}>0$ was described in \cite{dHPP} where it was found that the instanton solution signals an instability of EAdS against deformations by mixed boundary conditions that can be described via a Coleman-de Luccia scenario. The decay rate calculation was presented in \cite{dHPP,Barbon}.

The result of the corresponding calculation \eq{HHwf} for the de Sitter case is:
\bea\label{finalS4}
-\log\Psi_{\tn{HH}}=S^{\tn{on-shell}}_{S_-^4}&=&-{4\pi^2\ell^2_{S^4}\over\k^2}-{\l_{\tn{dS}}\,\pi^2\ell_{S^4}^4\over12}{2a_5-a_0\over3a_5^3(a_5-a_0)^2}-\pi^2\ell^2_{S^4}\,{a_0\over a_5^3}\nn
&=&-{4\pi^2\ell^2_{S^4}\over\k^2}+{2\pi^2\ell^2_{S^4}\over3a_5^3}~{a_5^2-a_5a_0+a_0^2\over a_5-a_0}~,
\eea
and again we used the condition for the moduli space \eq{modulispacesdS}. If we take $a_0=0$, we get $a_5^2=-{\l_{S^4}\over12}\,\ell_{\tn{dS}}^2$ which can only be the if case $\l_{S^4}<0$. In that case, the on-shell value of the action is:
\bea
\label{zerocasedS}
S^{\tn{on-shell}}_{S_-^4}=-{4\pi^2\ell^2_{S^4}\over\k^2}-{8\pi^2\over\l_{S^4}}~,
\eea
which after analytic continuation $\ell_{S^4}\mapsto -i\ell_{\tn{AdS}}$, $\l_{S^4}\mapsto\l_{\tn{AdS}}$ agrees with \eq{zerocase}, as it should. Notice that this is negative for $\lambda_{S^4}<-2\kappa^2/\ell_{S^4}^2$ and vanishes for $\lambda_{S^4}=-2\kappa^2/\ell_{S^4}^2$, which coincides with  the critical value arising in the consistent truncation of ${\cal N}=8$ sugra, after the analytic continuation of the dS$_4$ radius.

Comparing \eq{finalEAdS} and \eq{finalS4} we see that the first, pure gravity, term matches under the analytic continuation $\ell_{\tn{AdS}}\rightarrow i\,\ell_{\tn{dS}}$. This is Maldacena's result in \cite{Maldacena2}. The result is non-trivial in part because the EAdS$_4$ needs to be regularized and renormalized, whereas the $S_-^4$ calculation of the HH wave function is completely finite. In order to match the matter contributions, however, in the second term we need to analytically continue the couplings as well. This analytic continuation from EAdS$_4$ to $S_-^4$ is an invertible map $\g$:
\bea\label{ancont}
\g(\ell_{\tn{AdS}})&=&i\,\ell_{\tn{dS}}\nn
\g(b_0)&=&i\,a_5\nn
\g(b_5)&=&-i\,a_0\nn
\g(b_i)&=&ia_i~.
\eea
It follows that the coupling constant does not change, i.e.~$\g(\l_{\tn{AdS}})=\l_{\tn{dS}}$. Notice that the fact that the moduli have to be analytically continued is a consequence of the prescription to analytically continue the EAdS$_4$ radius of curvature. Then the two expressions exactly match.  

In contrast to \cite{Maldacena2,Anninos:2014hia}, we did not need to write absolute value bars around the HH wave function \eq{finalS4} because this is the full semi-classical result \eq{HHwf} which is real in the Euclidean. This result can be directly compared after analytic continuation, as we have seen, to the Euclidean AdS/CFT partition function because the latter is finite---we took into account the correct counterterms, thereby rendering a result that can be precisely matched without the need to take the real part. 

The complexification of the moduli can be understood from the fact that they are dimensionful quantities, to be measured in units of the radius. Defining dimensionless moduli $y_I=a_I/\ell_{\tn{dS}}$, $z_I=b_I/\ell_{\tn{AdS}}$, $I=0,\ldots,5$, we find the the moduli spaces can be represented as:
\bea
-z_0^2+z_5^2+z_i^2&=&{\l_{\tn{AdS}}\over12}~~(i=1,\ldots,4)\nn
y_0^2-y_5^2+y_i^2&=&{\l_{\tn{dS}}\over12}~.
\eea
These are all real quantities on both sides. The moduli space is O(1,5) invariant:
\bea
\eta^{IJ}y_I\,y_J={\l_{\tn{dS}}\over12}~,~~I,J=0,\ldots,5~,
\eea
with $\eta^{IJ}$ the O(1,5) Minkowski metric. The analytic continuation is then simply an SO(1,5) map of the moduli space onto itself:
\bea
z_I=\e_I{}^Jy_J~;~~~\e=\d\left(\begin{array}{ccc}0&1&\\-1&0&\\&&\mathbb{1}_{4\times4}\end{array}\right)\,,\,\,\,\delta=\pm 1.
\eea

We note here that in order to enforce the boundary conditions \eq{bdydef} from the bulk equations of motion, one needs to add a further term to the action: 
\bea\label{bdydefaction}
S_{\tn{bdy def}} = -{b_5\ell_{\tn{AdS}}^2\over3} \int\dd\O_3~\F_{(0)}^3(\O) =-{2\pi^2\ell^2_{\tn{AdS}}b_5\over3b_0^3}~.
\eea
This of course agrees, after the analytic continuation \eq{ancont}, with the term one gets in the dS$_4$ case, $-{2\pi^2\ell^2_{\tn{dS}}\,a_0\over3a_5^3}$. In the presence of these terms, the on-shell actions are obtained simply by adding (\ref{bdydefaction}) and the corresponding dS$_4$ result to \eq{finalEAdS}-\eq{finalS4}.

\section{Geometric interpretation}

The 3-sphere partition function $Z_{S^3}$ for a three-dimensional CFT measures its number of degrees of freedom. Moreover, it has been argued that for unitary CFTs the corresponding free energy is given (in suitably chosen units) by
\be
F=-\log |Z_{S^3}|\,,
\ee
which is positive and satisfies an $F$-theorem, namely it  decreases along RG flows from the UV to the IR \cite{Jafferis,JKPS}. Holographically, the partition function is usually calculated using the bulk gravitational action on EAdS$_4$ with all other matter fields set to zero. The result is the first term in (\ref{zerocase}) and it is proportional to the dimensionless ratio $\ell^2_{\tn{AdS}}/\kappa^2$.  One may then wonder what the physical interpretation is of the second term in \eq{zerocase}, which corresponds to the contribution of the bulk scalar fields. Notice that for $\lambda_{\tn{AdS}}<0$ this term is also positive. A similar question may be asked for the result (\ref{zerocasedS}), namely whether this can also be interpreted as a partition function of a CFT$_3$ on a 3-sphere. Since (\ref{zerocasedS}) is not always positive, such a CFT$_3$ need not be unitary.

To this end, we will point out that the instanton contributions in (\ref{zerocase}) and (\ref{zerocasedS}) also arise from bulk gravitational actions, and hence can be interpreted as $F$-functions on the 3-sphere. This follows \cite{StoQ} from the well-known representation of the conformal factor of a metric as a scalar field with quartic self interaction. Indeed, consider the conformally related metrics
\be
\label{ghmetrics}
g_{\m\n}=\Omega^{-2}\,h_{\m\n}
\ee
One can then show\footnote{To obtain the rhs of (\ref{EHconfscal}) we used the fact that the trace of the extrinsic curvature for the conformally related metrics (\ref{ghmetrics}) are related as
$$
\label{Khg}
K[h]=\frac{1}{\Omega}\,K[g]+\frac{3}{\Omega^2}\,n^\m\partial_\m\Omega\,,
$$
with $n^\m$ the outward normal to the boundary in the metric $g_{\m\n}$.} that
\bea
\label{EHconfscal}
-I_h&:=&\frac{1}{2\kappa^2}\int_{M_1} \dd^4x\,\sqrt{h}\left(R[h]-2\Lambda_h\right)  +\frac{1}{2\kappa^2}\int_{\partial M_1} \dd^3x\,\sqrt{\gamma_{h}}~2K[h]\\
&=& \int_{M_2} \dd^4 x\,\sqrt{g}\left(\frac{1}{2}\,(\pa_\m\f)^2 +\frac{R[g]}{12}\,\phi^2+\frac{\lambda}{4!}\,\phi^4\right)+\frac{1}{6}\int_{\partial M_2}\dd^3x\,\sqrt{\gamma_g}~K[g]\,\phi^2
=: I_{(g,\phi)}\nonumber
\eea
where $\gamma_h$ and $\gamma_g$ are the induced metrics of $h_{\m\n}$ and $g_{\m\n}$ on the boundary $\partial M_1,\pa M_2$ and we have defined
\be
\label{rescalings}
\phi :=\sqrt{\frac{6}{\kappa^2}}~\Omega\,,\,\,\,\,~~~\lambda :=-\frac{2\kappa^2}{3}\,\Lambda_h\,.
\ee

The critical value for $\lambda$ mentioned in the previous section arises when we relate the scalar action to a gravitational action. The second line in (\ref{EHconfscal}) coincides with the matter part of the action (\ref{action}), and from the above it equals {\it minus} a gravitational action including the GH term.  other words, 

The crucial observation now is that the instanton solutions $\phi_{\tn{inst}}$ on either $S^4_-$ and EAdS$_4$, with the moduli set to specific values, correspond exactly to the conformal factor relating the two metrics.  Hence, the on-shell action of instantons on EAdS$_4$ corresponds to the volume of (half) $S^4$ and conversely, the on-shell action of instantons on the half $S^4$ corresponds to the volume of EAdS$_4$, where the instanton deformation parameter $b$ (which corresponds to $a_0$ in the previous section) regulates the volume of EAdS$_4$. It is also crucial to point out that in evaluating the on-shell values of the gravitational actions, the boundary GH term does not contribute to the finite part. This is true for AdS$_4$, but it is also true for $S^4$, since in this case the extrinsic curvature vanishes. 

Let us see how this arises. On the half $S^4$ with curvature radius $\ell$ and metric
\be
\label{S4/2}
\dd s^2=\frac{4}{\left(1+\frac{\rho^2}{\ell^2}\right)^2}\left(\dd\rho^2+\rho^2\dd\Omega_3^2\right)
\,,
\ee
the instanton solutions are given by
\be
\label{instS4}
\phi_{\tn{inst}}(\rho)=\pm\sqrt{\frac{12}{\lambda_{S^4}}}\frac{1}{b}\frac{1+\frac{\rho^2}{\ell^2}}{1-\frac{\rho^2}{b^2}}
\ee
where $\lambda_{S^4}>0$. It is important to note the presence of the instanton  modulus $b$ which is in principle unrelated to $\ell$.  In particular, since the range of the radial coordinate is $\rho\in [0,\ell)$,  if we consider $b>\ell$ the solution is everywhere regular.

 Next we notice  that half $S^4$ and EAdS$_4$ are conformally related metrics. In particular
 \be
 \label{S4EAdS4}
 \dd s^2=\frac{4}{\left(1+\frac{\rho^2}{\ell^2}\right)^2}\left(\dd\rho^2+\rho^2\dd\Omega_3^2\right) =\left(\frac{1-\frac{\rho^2}{b^2}}{1+\frac{\rho^2}{\ell^2}}\right)^2\frac{4}{\left(1-\frac{\rho^2}{b^2}\right)^2}\left(\dd\rho^2+\rho^2\dd\Omega_3^2\right)\,,
 \ee
where on the right EAdS$_4$ has a generally different radius $b$ which is set equal to the instanton modulus. Hence, the calculation of the on-shell action for instantons on half $S^4$---including an appropriate boundary term as in (\ref{EHconfscal})---boils down to the calculation of the {\it regularized} volume  of  EAdS$_4$. Notice that it is the presence of the instanton modulus $b>\ell$ that gives rise to a particular regularization of the volume of the EAdS$_4$ space with radius $\ell$. 

Explicitly,  we obtain
\be
\label{IhS4}
I_{(g,\phi)}^{\tn{on-shell}}\left(S^4_-\right) = \frac{\Lambda_h}{\kappa^2}\int \dd^4x\,\sqrt{h}=\frac{8\pi^2 b^2}{\kappa^2}\frac{\alpha^4(\alpha^2-3)}{(1-\alpha^2)^3}=\frac{16\pi^2}{\lambda_{S^4}}\frac{\alpha^4(\alpha^2-3)}{(1-\alpha^2)^3}\,,\,\,\,~~\alpha=\frac{\ell}{b}<1\,.
\ee
This diverges as $\alpha\rightarrow 1$, which we shall interpret presently. The finite part of (\ref{IhS4}) for $\alpha\rightarrow 1$ coincides with {\it minus} the first term in (\ref{zerocase}), as it should. 

Next we consider the instanton solutions on EAdS$_4$ with radius $\ell$. These have the form
\be 
\label{EAdSinst}
\phi_{\tn{inst}}(\rho)=\pm\sqrt{\frac{12}{-\lambda_{\tn{AdS}}}}\frac{1}{b}\frac{1-\frac{\rho^2}{\ell^2}}{1+\frac{\rho^2}{b^2}}\,.
\ee
with $\lambda_{\tn{AdS}}<0$ and $b$ the modulus which we take here again to be $b>\ell$. The on-shell action would give part of the volume of $S^4$ with  radius $b$. We find
\be
\label{IhEAdS4}
I_h^{\tn{on-shell}}\left(\mbox{EAdS}_4\right)=\frac{8\pi^2 b^2}{\kappa^2}\frac{\alpha^4(\alpha^2+3)}{(\alpha^2+1)^3}=-\frac{16\pi^2}{\lambda_{\tn{AdS}}}\frac{\alpha^4(\alpha^2+3)}{(\alpha^2+1)^3}~,
\ee
where here $\lambda_{\tn{AdS}}<0$. For $\alpha=1$ this is just {\it minus} the first term in (\ref{zerocasedS}), as it should. 

The association of the on-shell instanton actions with {\it minus} the EH action in (\ref{EHconfscal}) points towards a remarkable correspondence. Firstly, it is not hard to see that (\ref{IhS4}) and (\ref{IhEAdS4}) are mapped to each other by the transformation $b\mapsto i\,b$. This is natural, if we recall that  $b$ is the radius of both the $S^4$ and EAdS$_4$ of the associated EH actions. Next, we should remember that the AdS$_4$ result (\ref{zerocase}) gives (minus the logarithm) of the partition function of a three dimensional CFT on $S^3$ as a sum to two terms. The first term is the usual holographic $F$-function that is proportional to dimensionless ratio $\ell^2/\kappa^2$. What we have shown here is that the second term too has a geometric origin, since it also comes from a gravitational action which is nevertheless in principle unrelated to the AdS$_4$ action. Namely, if we identify the coupling $\lambda \sim \kappa'^2/b^2$, the second term is the on-shell value of a gravitational action with Newton's constant $\kappa'$ and {\it positive} cosmological constant, in other words it gives the volume of an $S^3$ with radius $b$.  Moreover, the ratio of the two radii, $\alpha=\ell/b$, which also has a geometric origin, can be viewed as a natural deformation parameter, running from $b=\infty~(\alpha =0)$ to $b=0~(\alpha=\infty)$. It is worth noting then that such a deformation is monotonic and according to the conjectured $F$-theorem takes the boundary CFT towards the UV.  The case of dS$_4$ is exactly the reverse; here it is the scalar instanton part that has a natural holographic interpretation as a partition function of a CFT on $S^3$, while the gravitational part is geometric. Here too, the parameter $\alpha$ deforms monotonically the (non unitary) boundary theory.

We notice here that the instanton \eq{instS4} is of the type \eq{solutions} under the following identification:
\bea
b^2 = \ell^2_{\tn{dS}}\,\frac{a_0-a_5}{a_0+a_5},
\eea
as is easy to verify doing the coordinate transformation. We see that the analytic continuation $b\mapsto ib$ indeed corresponds to $\ell_{\tn{dS}}\mapsto-i\,\ell_{\tn{AdS}}$. 

The limit $\a\rightarrow1$ of the above action, in which the instanton deformation parameter approaches the curvature radius, corresponds to the limit $a_5^2\rightarrow0^+$ and $a_0^2\rightarrow{\l_{S^4}\over12}\ell_{\tn{dS}}^2{}^+$ which is the critical value at which the moduli space shrinks to zero curvature radius. In this limit, the instanton part of the action \eq{finalS4} diverges as $1/a_5^3$, which is precisely the divergence of the EAdS$_4$ volume\footnote{This divergence can be removed by adding boundary terms in a way similar to what was done in section 2, however our aim here is to {\it exhibit} the form of this divergence and how it relates to the on-shell value of the instanton.}. Thus, the instanton is computing the EAdS$_4$ volume, and the boundary deformation parameter $a_0$ regulates this volume. The critical value of the deformation parameter $a_0^2={\l_{S^4}\over12}\ell_{\tn{dS}}^2$, for which we get the correct divergence, corresponds, via the analytic continuation \eq{ancont}, to the critical value at which the dual boundary theory becomes unstable.

\section{Discussion and conclusions}

In this note we have tested the HHM proposal in the case of scalar instantons. We have calculated the on-shell action for instantons on half-$S^4$, which yields the late-time HH state, and compared it with the on-shell action for instantons on EAdS$_4$. The results match under the HHM prescription of analytically continuing the curvature radii. Additionally, we have found that it is also necessary to analytically continue the boundary condition, which corresponds to a marginal triple trace deformation. This provides new evidence that the HHM proposal works for exact, but non-trivial configurations.

Our goal was to check the HHM proposal but along the way we got a number of results which we believe are relevant for the HS proposals in de Sitter space, in particular for the $\mbox{Sp}(N)$ vector model that has been conjectured to be dual to it. On the EAdS side, the instanton modulus $b_5$ corresponds to the coefficient of a marginal triple trace deformation \eq{bdydefaction} for an operator of dimension 1. Under the analytic continuation to dS space this is again a triple trace deformation, and the free energy has been computed. The free energy as a function of the boundary deformation parameter presents zeroes which usually signal an instability of the theory. The presence of instabilities seems to be connected to the fact that our free energy result (\ref{zerocasedS}) may be negative, hence it corresponds to a non-unitary CFT$_3$.  Similar behaviour has been found in \cite{Anninos:2014hia}. It would also be interesting to work out the relationship of this result with holographic stochastic quantisation \cite{StoQ,Jatkar}.

We also found an interesting geometric realization of the same computation, in which the $S^4$ instantons (for particular values of the moduli) are seen to compute the regularized volume of EAdS$_4$, and the EAdS$_4$ instantons are seen to compute the volume of the four-sphere. The regulator of the EAdS$_4$ volume is $a_0$, with the divergence appearing precisely for the critical value $a_0^2={\lambda_{S^4}\over12}\,\ell_{\tn{dS}}^2$. This might imply that a sector of the $\mbox{Sp}(N)$ model with this marginal deformation is dual to a pure gravitational theory with no scalars, and hence signal a duality between $\mbox{Sp}(N)$ models with different values of the deformation parameter. This is reminiscent of similar scenarios as e.g.~the dualities in \cite{SdH,SdHPG}.

As stressed in \cite{SdHTP,SdHTP2}, instantons of the type found here seem to have special holographic properties because they parametrize different Weyl vacua of the theory. The exact bulk action can be calculated and compared to the boundary effective potential \cite{dHPP,Iannis}. It was pointed out in \cite{SdHTP} that the boundary values of bulk instantons are also solutions of the equations of motion of a three-dimensional conformally coupled scalar field theory on the boundary, which was conjectured to be the effective action for an operator of dimension 1. In \cite{dHPP,Iannis} it was found that this action in fact agrees with the effective action near the critical point, which can be calculated by different methods. Given the robust structure of the instanton solutions it is unlikely that this is a mere coincidence. Near the critical point, as we have seen, the instantons in fact just calculate the volume. We point out here that a similar boundary effective action description applies to the de Sitter instantons as well. 

Finally, we wish to point out that our instanton solutions are intimately related to the  $\mbox{SO}(4)$ and $\mbox{SO}(3,1)$ invariant solutions of 4-dimensional HS theory found in \cite{SS}. The latter are solutions with all HS gauge fields switched off, except of the metric and the conformally coupled scalar, and they are also related \cite{dHPP} to a consistent truncation of ${\cal N}=8$ gauge supergravity down to a single scalar of the $\mbox{SO}(8)$ group. In that sense, our results should also provide the partition functions of both the above theories, at the scalar instanton vacua.  We leave a more detailed analysis of this intriguing point of view for future work.

\section*{Acknowledgements}\addcontentsline{toc}{section}{Acknowledgements}

We thank D.\ Anninos, J.\ de Boer, K.\ Skenderis, and P.\ Sundell for useful discussions, communications  and comments on the manuscript. SdH thanks the Institute for Theoretical Physics of the Aristotle University of Thessaloniki for the kind hospitality during the final stages of this work. The work of SdH was partially supported by EU-COST action ``The String Theory Universe'', STSM-MP1210-16939. The work of A.~C.~Petkou is partially supported by the research grant ``ARISTEIA II", 3337 ``Aspects of three-dimensional CFTs", by the Greek General Secretariat of Research and Technology, and also by the CreteHEPCosmo-228644 grant. 

\appendix

\section{Explicit bulk solutions and holography}
The near boundary behaviour of a conformally coupled scalar field $\phi(z,\vec{x}) $ on a fixed EAdS$_4$ background is as
\be
\label{confscal}
\phi(z,\vec{x}) \rightarrow z\,\phi_{(0)}(\vec{x}) +z^2\phi_{(1)}(\vec{x})+\cdots
\ee
One can then calculate the renormalized bulk on-shell action $I_{\tn{on-shell}}[\phi_{(0)}]$ as a functional of the boundary conditions $\phi_{(0)}$. This is interpreted as (minus) the generating functional for connected correlation functions of the boundary operator ${\cal O}$ as:
\be
\label{IW}
I_{\tn{on-shell}}[\phi_{(0)}] = -W[\phi_{(0)}]\,,\,\,\,\,~~~~\frac{\delta W[\phi_{(0)}]}{\delta\phi_{(0)}}=\langle {\cal O}\rangle_{\phi_{(0)}}~.
\ee
Defining then the Legendre transform $\Gamma[{\cal A}]$ as
\be
\label{Legendre}
W[\phi_{(0)}]=\Gamma[{\cal A}]+\int {\cal A}~\phi_{(0)}~,
\ee
we have
\be
\frac{\delta\Gamma[{\cal A}]}{\delta{\cal A}}=-\phi_{(0)}\,,\,\,\,~~~~~{\cal A}=\langle {\cal O}\rangle_{\phi_{(0)}}~,
\ee
which shows that $\Gamma[{\cal A}]$ is the effective action of the boundary theory. Knowledge of $\Gamma[{\cal A}]$ allows us to study non-trivial vacua of the boundary theory, which are then given by the set of equations
\be
\label{ntvacua}
\frac{\delta \Gamma[{\cal A]}}{\delta {\cal A}}\Biggl|_{{\cal A}={\cal A}_*}=0\,,\,\,\,\,\,~~~~\frac{\delta W[\phi_{(0)}]}{\delta\phi_{(0)}}\Biggl|_{\phi_{(0)}=0}=\langle {\cal O}_*\rangle \equiv {\cal A}_*\neq 0~.
\ee

Now, suppose that we impose mixed boundary conditions by adding a boundary term of the form $f(\phi_{(0)})$ to the bulk action. We then obtain for the variation of the on-shell bulk action:
\be
\label{variationOS}
\delta I_{\tn{on-shell}}[\phi_{(0)}]=\int \delta\phi_{(0)}\,(\phi_{(1)}-f'(\phi_{(0)}))~.
\ee
This means that we are rendering the bulk on-shell action stationary for solutions of the bulk equations of motion satisfying
\be
\label{Neumanbc}
\phi_{(1)}=f'(\phi_{(0)})~.
\ee
This in turn implies that 
\be
\label{dtheory}
\frac{\delta W[\phi_{(0)}]}{\delta\phi_{(0)}}\Biggl|_{f'[\phi_{(0)}]=\phi_{(1)}}=0~.
\ee
Therefore, we could have interpreted $W[\phi_{(0)}]$ as an effective action of a {\it dual} boundary theory that has a non-trivial vacuum structure. In fact, we can write \cite{dHPP} 
\be
\label{dualea}
W[\phi_{(0)}]\equiv \tilde{\Gamma}[-\phi_{(0)}]\,,\,\,\,\,~~~ ~~\tilde{W}[-\sigma]=\tilde{\Gamma}[-\phi_{(0)}]+\int\phi_{(0)}\,\sigma
\ee
such that 
\be
\label{dualgenf}
\frac{\delta \tilde{W}[-\sigma]}{\delta\sigma}\Biggl|_{\sigma=0}=-\langle\tilde{\cal O}_*\rangle\neq 0
\ee
yields the non-trivial vacuum  expectation value for the operator $\tilde{\cal O}$ in the dual boundary theory.

The partition function is given by 
\be
\label{pf}
Z[J]=\int {\cal D}\varphi~e^{-S[\varphi]+\int J{\cal O}}=e^{W[J]}=e^{\Gamma[{\cal A}]+\int J{\cal A}}
\ee
If the theory possesses non-trivial vacua (\ref{ntvacua}), then setting the external source $J=0$ we obtain
\be
Z[J=0]\equiv Z_0=e^{\Gamma[{\cal A}_*]}=e^{-S[\varphi_*]+\cdots}
\ee
Hence, the leading part of the effective action gives the partition function.

From the AdS/CFT point of view, given an exact {\it particular} solution of the bulk equations of motion, namely a solution with a given form $\phi_{(0)}\equiv \bar{\phi}_{(0)}$ in (\ref{confscal}), then we would be able to evaluate $W[\bar{\phi}_{(0)}]$. According to (\ref{dualea}) this would give us the partition function $\tilde{Z}_0$ of the {\it dual} boundary theory as
\be
\label{pfdual}
W[\bar{\phi}_{(0)}]=\tilde{\Gamma}[-\bar{\phi}_{(0)}]=\ln\tilde{Z}_0
\ee
On the other hand, given $W[\bar{\phi}_{(0)}]$ we are able using (\ref{Legendre}) to obtain the value of the effective action of the boundary theory at its non-trivial vacuum ${\cal A}_*$, as
\be
\label{pfnormal}
\Gamma[{\cal A}_*]=W[\bar{\phi}_{(0)}]-\int {\cal A}_*\bar{\phi}_{(0)}=\ln Z_0
\ee

\end{document}